# Tuning Dipolar and Multipolar Resonances of Chiral Silicon Nanostructures for Control of Near field Superchirality.

Dominic J.P. Koyroytsaltis-McQuire[1*], Rahul Kumar[1], Tamas Javorfi[2], Giuliano Siligardi[2], Nikolaj Gadegaard[3], Malcolm Kadodwala[1*]

[1] School of Chemistry, University of Glasgow, Glasgow, G12 8QQ, UK

[2] Diamond Light Source Ltd., Harwell Science and Innovation Campus, Didcot, United Kingdom

[3] School of Engineering, Rankine Building, University of Glasgow, Glasgow G12 8LT, UK

* Corresponding Authors

E-mail: Malcolm.kadodwala@glasgow.ac.uk

E-mail: d.koyroytsaltis-mcquire.1@research.gla.ac.uk

## Abstract

Chiral materials display a property called optical activity, which is the capability to interact differentially with left and right circularly polarised light. This leads to the ability to manipulate the polarisation state of light, which has a broad range of applications spanning from energy efficient displays to quantum technologies. Both synthesised and engineered chiral nanomaterials are exploited in such devices. The design strategy for optimising the optical activity of a chiral material is typically based on maximising a single parameter, the electric dipole – magnetic dipole response. Here we demonstrate an alternative approach of controlling optical activity by manipulating both the dipole and multipolar response of a nanomaterial. This provides an additional parameter for material design, affording greater flexibility. The exemplar systems used to illustrate the strategy are nanofabricated chiral silicon structures. The multipolar response of the structures, and hence their optical activity, can be controlled simply by varying their height. This phenomenon allows optical activity and the creation of so called superchiral fields, with enhanced asymmetries, to be controlled over a broader wavelength range, than is achievable with just the electric dipole – magnetic dipole response. This work adds to the material design toolbox providing a route to novel nanomaterials for optoelectronics and sensing applications.

# Introduction

The amplification of chiral light -matter interactions are key for the development of novel sensing technologies. This is due to the inherent lack of sensitivity in traditional methods such as circular dichroism (CD) used to study chiral (bio)molecules. Nanophotonic platforms are key to delivering these new technologies since they provide a mechanism for significantly enhancing chiral light- matter interactions. One property that is central to sensing applications is the ability to generate electromagnetic fields with chiral asymmetry exceeding that of circularly polarised light (CPL), sometimes referred to as superchiral light. A chiroptical response of a material is usually defined by the electric dipole ($E_D$) and magnetic dipole ($M_D$) cross terms. Optimising the $E_DM_D$ terms is the usual strategy for maximising the chiroptical response in both nanophotonic[3–5] and molecular systems[6,7]. Studies which exploit nanophotonics to amplify chiral light matter interactions have predominately focused on plasmonic nanostructures.[8] This emphasis on plasmonic platforms is primarily driven by the ease of fabrication. However, a key limitation of plasmonic systems is high ohmic losses, which may be sub-optimal for some sensing and optoelectronic applications. This has led to a greater consideration for all-dielectric nano/metamaterials, which are capable of a wide range of electromagnetic responses and suffer low ohmic losses from the visible to telecommunications range[9–11]. In the context of using nanophotonics for amplifying chiral light matter interactions, dielectric materials have another significant advantage. Through controlling the dimensions of dielectric structures, electric and magnetic dipole resonances can be made to spectrally overlap, enhancing the near and far field chiroptical properties. However, this can only be achieved in a relatively narrow wavelength window within the visible region of the spectrum. In the case of anisotropic chiral materials, higher order multipolar terms, which average to zero in isotropic media, can also contribute to the chiroptical response. These multipolar contributions to chiroptical effects can be very significant (at least comparable to the $E_DM_D$ response) for nano/metamaterials which have near fields with intrinsically large gradient.[12] The concept tested in this study is that the near and far field chiroptical properties of a silicon metamaterial can be controlled by manipulating the dipole and multipolar responses through modifying a single form factor parameter, the height, of the constituent nanostructures. We show through a combination of experiment and numerical simulation, that chiroptical behaviour, including the superchirality of the near fields, has a significant contribution from the $E_D$ and electric quadrupole $E_Q$ cross terms. Birefringence effects, arising from the lack of four-fold rotational symmetries of the metasurfaces, have been decoupled from the true chiroptical responses using Mueller matrix polarimetry.

## Metasurface Design

The metasurfaces used in this study are based on an S structure, **figure 1a -c**, which is chiral due to the presence of the underlying substrate breaking the horizontal mirror symmetry. The S structures possess $C_1$ symmetry due to the asymmetric spacing between the top and bottom arms with respect to the middle arm, with the arms having 20 nm less separation, **figure 1a.** The S structure was used to create four distinct metasurfaces based on square arrays with periodicities of 850 nm. The four arrays, shown in **figure 1b**, include left and right-handed (LH and RH) enantiomorphs and two racemic types (i.e., they contain equal numbers of LH and RH structures and thus for perfect structures would have no net optical activity). One type of racemic array, referred to as RA, contains alternating domains consisting of 19x19 sub-arrays of either LH or RH structures. The other racemic form, referred to as RS, contains alternating individual LH and RH structures, of which an AFM image is shown in **figure 1c**. Each metasurface was fabricated to occupy 0.25 mm$^2$ and contains 600x600 structures. The metasurfaces where fabricated with structures of four different heights, 160, 180, 210, and 240 nm (see supplementary material **S1**).

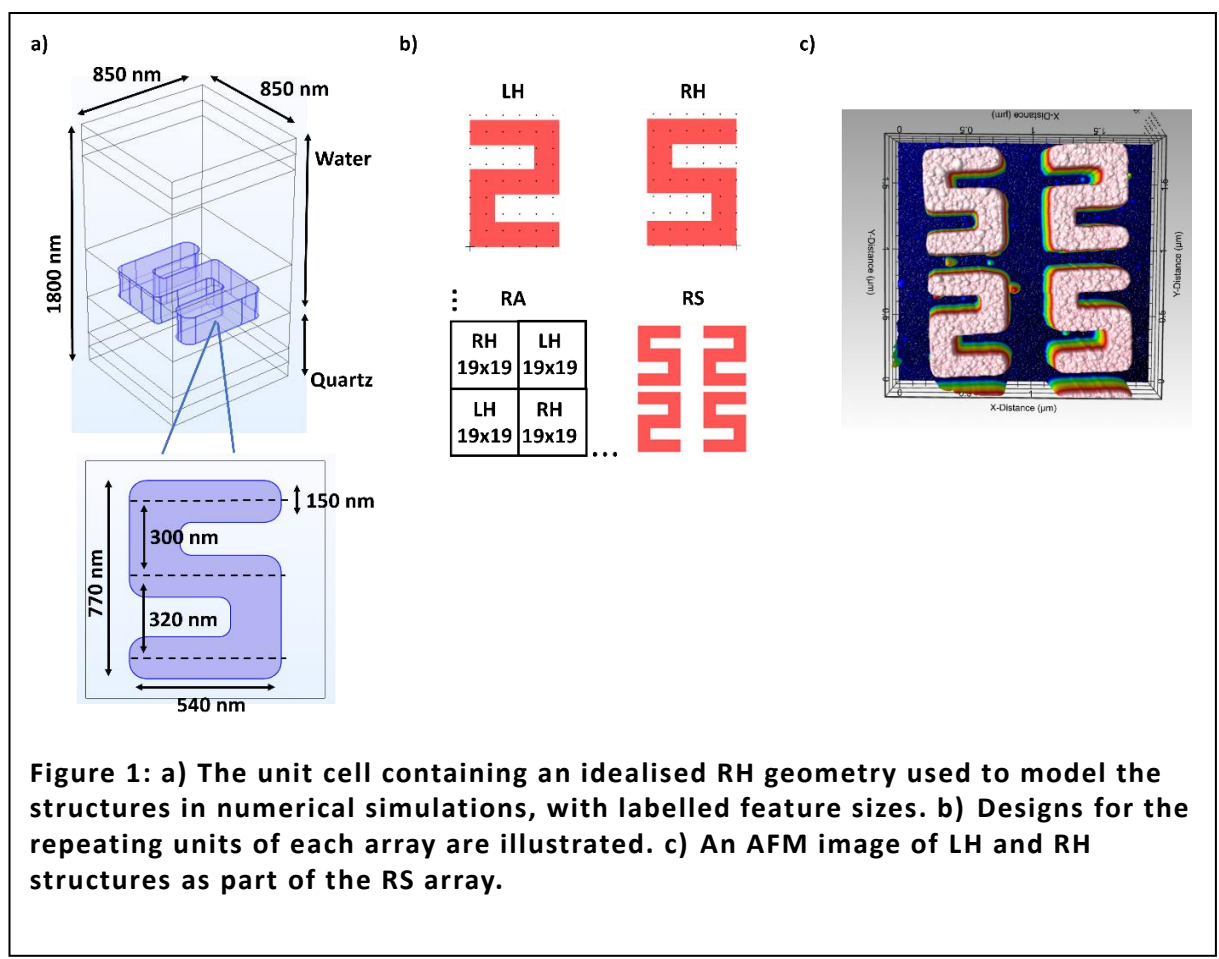

Figure 1: a) The unit cell containing an idealised RH geometry used to model the structures in numerical simulations, with labelled feature sizes. b) Designs for the repeating units of each array are illustrated. c) An AFM image of LH and RH structures as part of the RS array.

## Results and Discussion

### Reflectance Spectra

The light scattering behaviour of the metasurfaces were investigated via reflectance measurements, obtained relative to the substrate, for the four array types immersed in PBS buffer solution for each sample height, shown in **figure 2a-d**. Spectra were taken at two orthogonal incident polarisations of linearly polarised light (LPL) which are subsequently referred to and labelled as x- and y- polarisations.

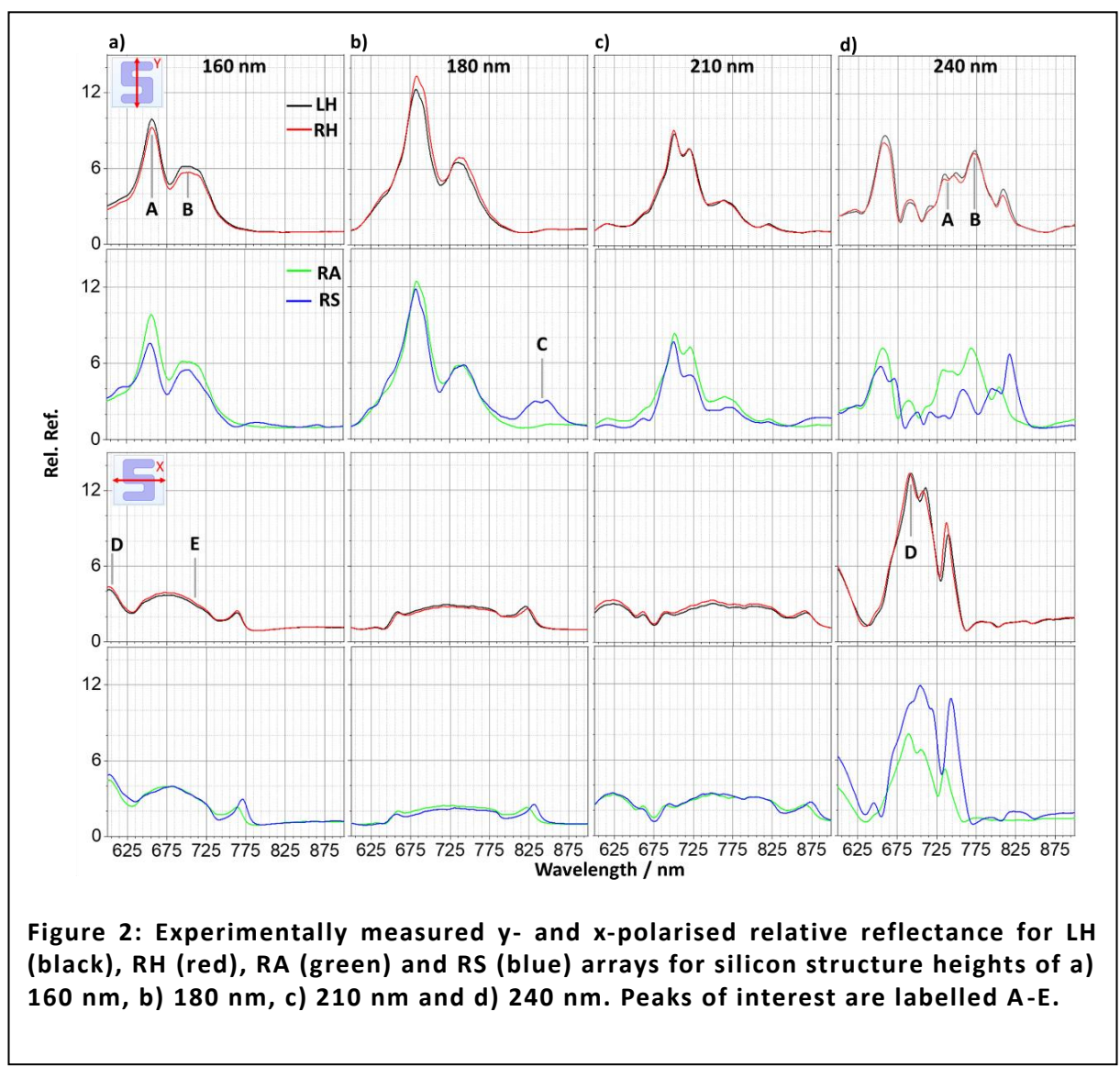

**Figure 2:** Experimentally measured y- and x-polarised relative reflectance for LH (black), RH (red), RA (green) and RS (blue) arrays for silicon structure heights of a) 160 nm, b) 180 nm, c) 210 nm and d) 240 nm. Peaks of interest are labelled A-E.

As would be expected, the spectra collected from the enantiomorphic pairs and the RA racemic structure are very similar. For the 160, 180 and 210 nm thick LH, RH and RA samples, **figure 2a-c,** the y-polarised spectra show a double peak, with two dominant components we have labelled A and B. There is a progressive red shift in reflectance features as the silicon height increases. For a thickness of 210 nm the A component undergoes a split. At a thickness of 240 nm there is an abrupt change in spectral profile, **figure 2d,** which is more significant than the progressive change observed between 160 to 210 nm. There is an appearance of a single peak close to 660 nm and a broad feature between 725 nm to 825 nm that contains multiple components.

The x-polarised spectra for LH, RH and RA substrates with heights 160 – 210 nm display broad and relatively weak reflectance, in contrast to intense resonances in the equivalent y-polarisation data. The general shape of the spectra (a broad peak situated between two individual peaks at 605 nm and 765 nm for the 160 nm sample) progressively red shifts as with the x-polarisation data. Once again, at a height of 240 nm the spectra undergo a significant change in form, with a large increase in reflectance intensity between 650 nm and 775 nm. The red shifting of resonances with increasing nanostructure height is analogous to the previously reported effects of increasing size on silicon nanoparticles[13–15]. It is consistent with behaviour expected from Mie theory, where resonances are directly proportional to the diameter of the scatterer.[16,17]

The reflectance response of the RS metasurfaces are qualitatively like that of the other three. There is a very strong similarity between y-polarised RS spectra for 160 to 210 nm thick substrates and those from the other arrays. However, the RS arrays have an additional resonance at ~ 850 nm, which correlates to the periodicity of the structure, and has subsequently been labelled C. There is a significantly greater difference in spectral line shape for the 240 nm thick substrate in the 650 to 850 nm range. The x-polarisation RS data has smaller differences in overall spectral line shape with those of the other arrays.

## Numerical Simulation of Reflectance Data

Numerical electromagnetic modelling has been used to provide insight into the light scattering properties of the silicon metasurfaces. Simulated reflectance spectra, relative to the substate, derived from modelling are shown in **figures 3a-d**. As expected, data for LH and RH enantiomorphs are identical (see supp. **S2**), so only the latter are shown. We have assumed that simulated data for the RA racemic structure will be the average of that obtained from the two separate enantiomorphic structures.

The simulated reflectance data for both x- and y-polarisation are in reasonable agreement with experimental measurements, both qualitatively replicating line shape and the abrupt change that occurs between 210 and 240 nm. However, the spectra display resonances which are significantly more intense, by a factor of ~20, than are observed experimentally. This in part can be attributed to structural and compositional heterogeneity (missing structures and chemical contamination which can lead to doping of the silicon and hence larger carrier densities) which are not accounted for in the model. A similar level of discrepancy between simulation and experiment has been observed previously in plasmonic nanoarrays.[2]

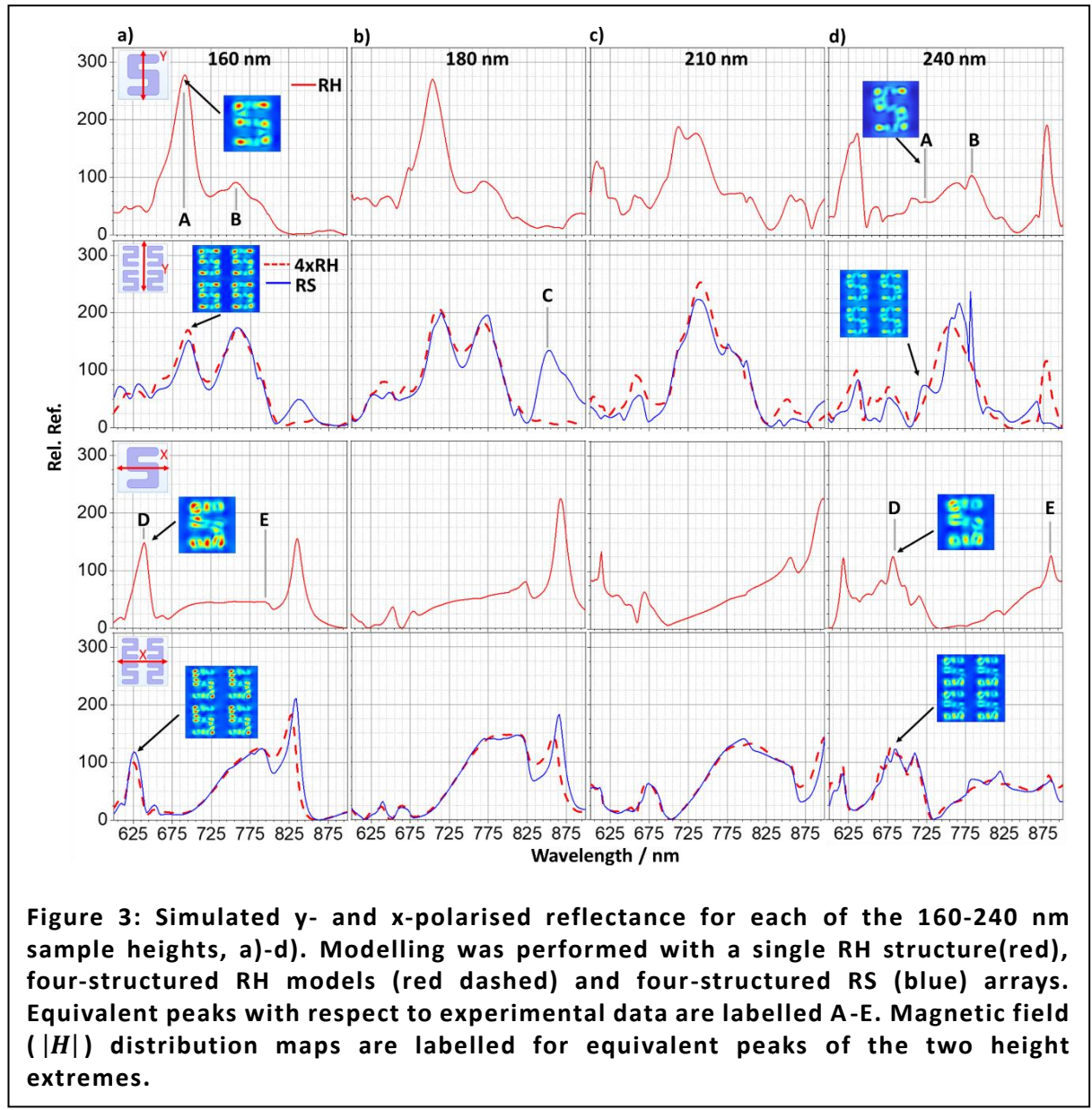

**Figure 3:** Simulated y- and x-polarised reflectance for each of the 160-240 nm sample heights, a)-d). Modelling was performed with a single RH structure(red), four-structured RH models (red dashed) and four-structured RS (blue) arrays. Equivalent peaks with respect to experimental data are labelled A-E. Magnetic field ($|H|$) distribution maps are labelled for equivalent peaks of the two height extremes.

The y-polarised modelling for the enantiomorphic pairs and RS array replicates the double peaks, A and B, for the 160 nm and 180 nm structures, as well as the peak splitting that occurs for peak A at 210 nm thickness. The 240 nm sample reproduces the single peak below 700nm and the broader range of peaks across the 700 – 800 nm range.

The x-polarised simulation for the enantiomorphic pairs and RS displays poorer agreement with experiment but do replicate the presence of two peaks either side of a broader region of intensity for the 160 – 210 nm samples. The simulations do however overestimate the intensity of the sharper peaks. There is further agreement for the 240 nm sample, in which a larger reflectance feature centred at ~ 700 nm is observed, although it is less intense compared to experiment. There is also a sloping effect that occurs as the sample height increases from 180 – 210 nm, where the reflectance intensity is greater at longer wavelengths.

To simulate the RS arrays requires using a model which has a unit cell of four S structures (see supp. **S3**), rather than a single structure used for enantiomorphic pair arrays. The larger model reduces the density of elements which degrades the accuracy of the simulations. In figure 3, our simulated spectra for RS are compared with RH spectra also derived from a four S containing unit cell. Although the four S models replicate the peaks observed for the single structure model, they have different relative intensities and a loss of feature resolution consistent with the expected decreased accuracy of the simulation. However, the spectra derived from the simulations using four structures for RH and RS are very similar, with the RS showing the additional feature (peak C) observed experimentally. Further analysis of the four S simulations reveal that the characteristic field distributions and intensities are in close agreement with those produced by using simulations based on single structures (see supp. **S4**).

A common feature of the simulated data is the red shifting of spectra with respect to experiment. This is attributed to the lack of morphological defects (e.g. surface deformities and wall sloping that stems from limitations in the fabrication process) in the idealised "perfect" structures used for the simulations, which has been observed previously.[18] This hypothesis is validated using structural models derived directly from AFM measurements, **(**see supp. **S5**). The use of a 'real' AFM model geometry produced spectra in better agreement with experiment. In particular, the AFM model produces spectra with resonances blue shifted relative to those generated with idealised models. Due to increased computational resource/time demands associated with the large increase in meshing requirements for AFM models, idealised structures have been used in this work.

## Numerical Simulation Field Plots

In addition to the simulated spectra, plots showing the spatial distribution of electric ($|E|$) and magnetic ($|H|$) fields have also been generated. The veracity of using data derived from idealised rather than real models is supported by the similarity of the field distributions produced by each for equivalent resonances. (see supp. **S5**). Only RH fields are shown as LH are mirror image equivalents (see supp. **S6**) The field plots for resonances A, B, D and E for the four silicon heights for RH (RA) structure are shown in **Figure 4**. For each resonance, similar field distributions are observed for the four heights, albeit with varying intensities. This validates the assignment of the features in the spectra to common resonances which red shift with increasing silicon thickness. The spatial distributions of $|E|$ and $|H|$ fields have a common characteristic. Specifically, regions with the most intense $|E|$ correspond to areas where $|H|$ is weakest, and *vice versa*. This behaviour is consistent with that of other silicon nanomaterials, in which the electric field pattern represents circulating electric currents (see supp. **S7**).[19] For all resonances, $|H|$ intensities are strongly localised inside the silicon nanostructure. Also, resonance A has the largest $|E|$ outside the silicon, specifically between the arms of the S nanostructure.

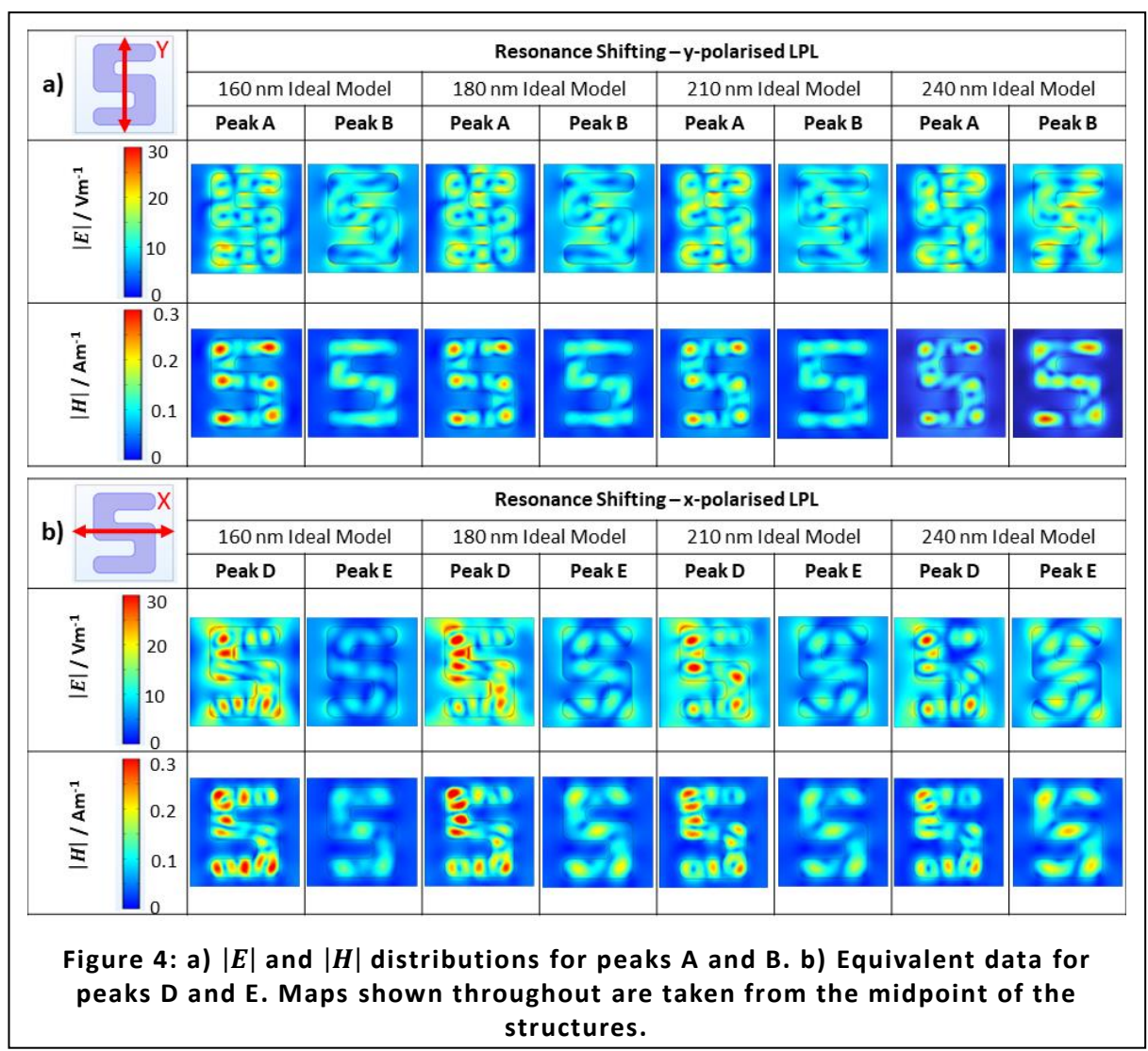

**Figure 4:** a) $|E|$ and $|H|$ distributions for peaks A and B. b) Equivalent data for peaks D and E. Maps shown throughout are taken from the midpoint of the structures.

Field distributions for resonance C, derived from models consisting of four-structured RH and RS arrays, are shown in **figure 5.** The $|E|$ field distributions for the RS array show strong coupling resonances that occur between y-axis-adjacent structures. This indicates that resonance C is associated with coupling between structures, which we term a periodic resonance. This is not observed in the enantiomorphic case, which can be attributed to a less preferential "anti-parallel" alignment of the arms of adjacent structures, limiting their ability to couple. The absence of periodic resonances in the thicker samples can be rationalised by considering a previous study performed on silicon nanoparticle arrays. It was found that the increasing height of the constituent elements also produces a red shift in their periodic resonance.[20] In the case of our samples, the periodic resonance therefore shifts out of the detectable range. The simulated periodic resonances are more intense than those observed experimentally, likely due to structural imperfections. As such, the periodic resonance that occurs for the 160 nm simulation at ~ 835 nm (which is inherently less intense than the 180 nm sample) is not observed experimentally.

## Multipolar Analysis

To provide greater insight into the microscopic origins of the resonances observed in light scattering, a multipolar analysis has been performed. Such an analysis provides a framework for the Mie description of light scattering from homogeneous spherical nanoparticles.[16] Mie theory expands the scattering of incident light as a series of multipolar contributions. Multipole moments are mathematical functions used to represent the distribution of electromagnetic charge and current within a material. The total scattering field of a structure is therefore a superposition of the fields generated by multipole moments such as the electric dipole ($E_D$), magnetic dipole ($M_D$), electric quadrupole ($E_Q$) and magnetic quadrupole ($M_Q$), considered here.[21] It is therefore possible to extract each of their contributions numerically through multipole decomposition, a technique which has been performed on a variety of different structures.[22–25] For illustrative purposes, we have performed such analyses on enantiomorphic pairs for silicon heights of 180 and 240 nm, for both y- (**figure 6**) and x- (**figure 7**) light polarisations.

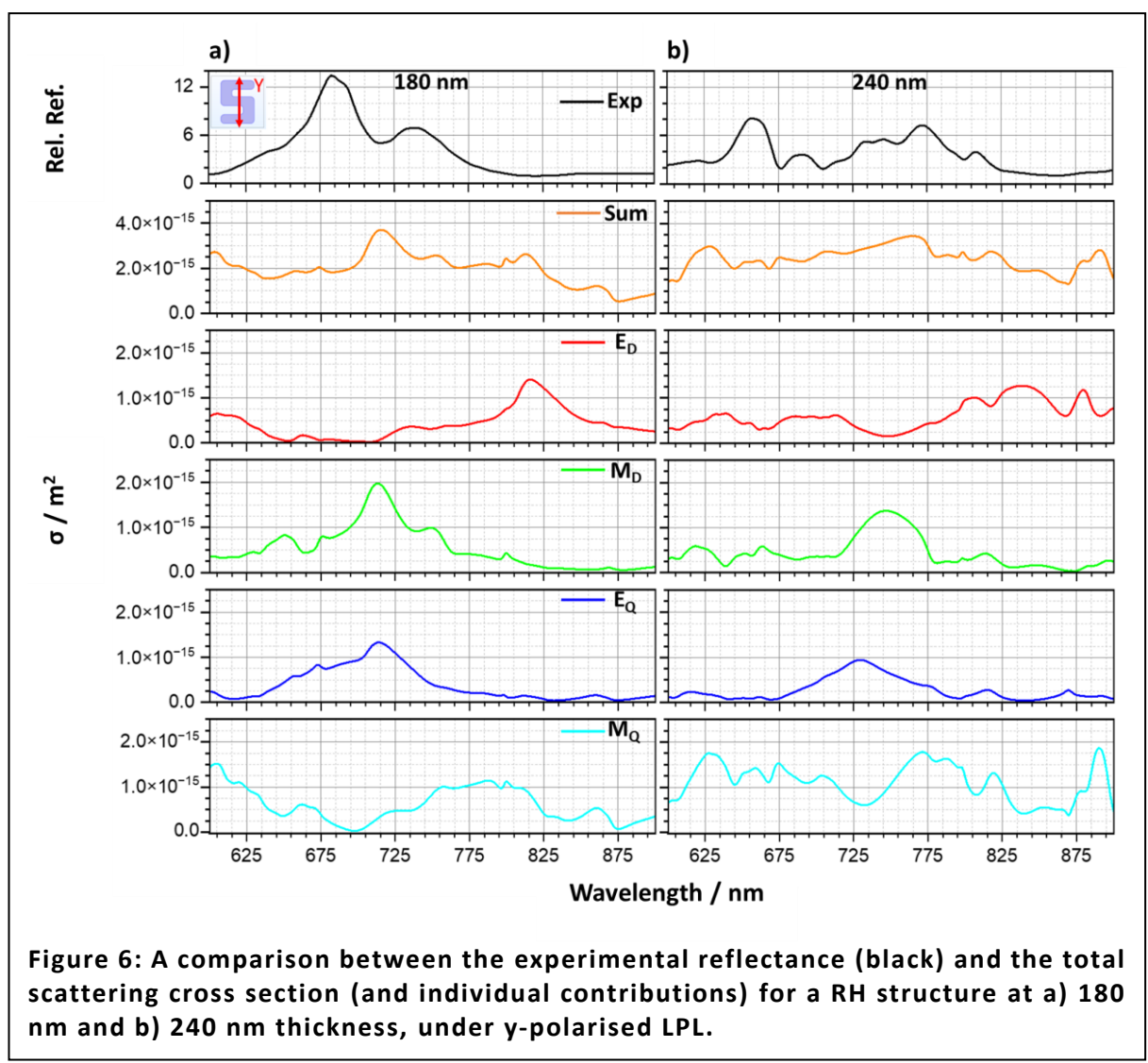

**Figure 6:** A comparison between the experimental reflectance (black) and the total scattering cross section (and individual contributions) for a RH structure at a) 180 nm and b) 240 nm thickness, under y-polarised LPL.

For the y-polarisation, the total scattering cross section shows qualitative agreement with the experimental reflectance for both sample heights in terms of peak shape. The dominant contribution towards peaks A and B is shown to be of $M_D$ character; peak A also has a significant $E_Q$ contribution. These multipole assignments agree with the field plots, where $|H|$ intensities (the $M_D$ resonance) are accompanied by circulating electric currents ($E_Q$). The multipolar resonances can be seen to red shift with increasing height of the silicon. There is also a significant change in the magnetic quadrupole contribution for the thicker sample.

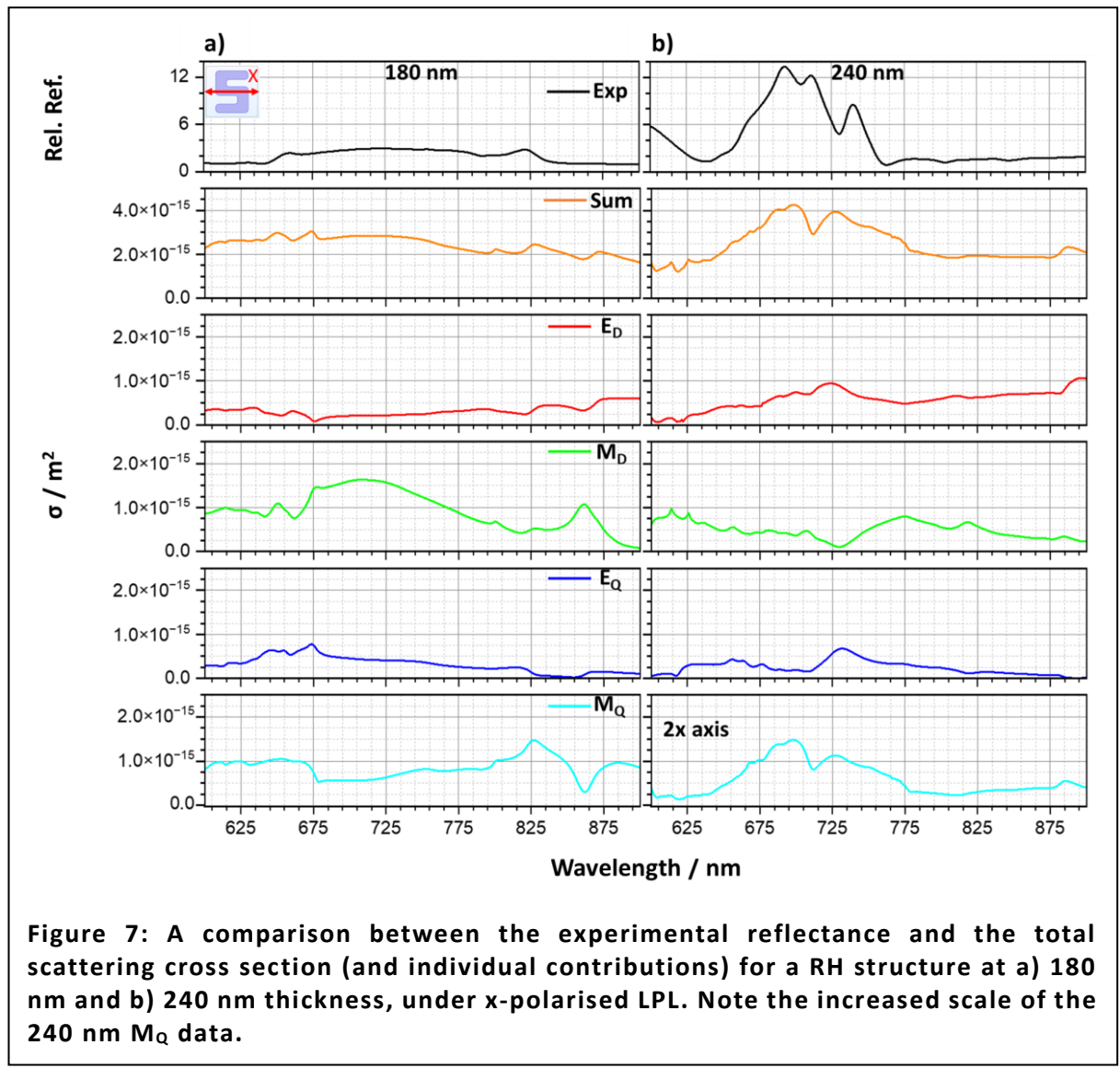

**Figure 7: A comparison between the experimental reflectance and the total scattering cross section (and individual contributions) for a RH structure at a) 180 nm and b) 240 nm thickness, under x-polarised LPL. Note the increased scale of the 240 nm $M_Q$ data.**

For the x-polarisation data, there is reasonable agreement between the experimental reflectance and the total scattering cross-section. The increase in reflectance for the 240 nm sample is repeated in the total scattering cross section, with a large increase contribution of the $M_Q$.

Overall, the total scattering calculated from the numerical simulations captures the general shapes of the reflectance spectra, although the relative peak intensities are not in as good agreement. This is likely a result of the scattering components being calculated in all directions, opposed to directional scattering (reflectance) measured in experiment and simulation. High index dielectric materials such as silicon produce a magnetic (first) dipolar resonance, in contrast to plasmonic materials, when excited by incident light.[17] This supports the dominant behaviour displayed by the structures being $M_D$ at lower sample thicknesses. The more dominant $M_Q$ contributions for the thicker sample can be rationalised by considering that the magnetic components of the resonances are parallel to the incident light excitation, therefore, as the silicon depth is increased, $M_Q$ is better supported across the height of the structures.[26]

The overall changes in the reflectance behaviour with increasing silicon thickness can be attributed to two sources. Those being a red shifting of resonances and significant increase in $M_Q$ contribution on increasing the thickness of the silicon to 240 nm.

## Optical Rotation from Stokes polarimetry

ORD spectra were obtained using a Stokes polarimeter for the samples for both incident polarisations, shown in **figure 8**.

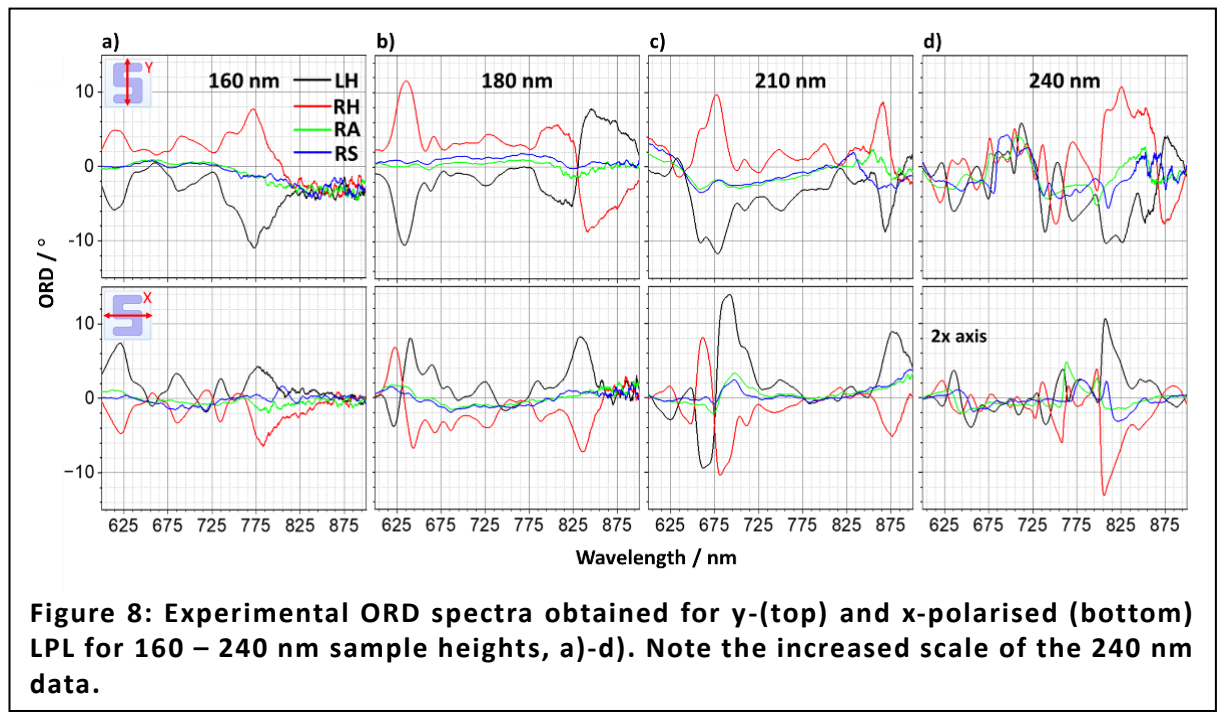

**Figure 8: Experimental ORD spectra obtained for y-(top) and x-polarised (bottom) LPL for 160 – 240 nm sample heights, a)-d). Note the increased scale of the 240 nm data.**

The line shapes change substantially with silicon thickness. Spectra from enantiomorphic arrays display approximately equal and opposite behaviour, whilst the racemic versions show no significant resonances, behaviour consistent with optical activity. However, significantly, the line shape of the spectra depends strongly on the incident polarisation of light. This is inconsistent with an optical active chiral response, clearly indicating the optical rotation is dominated by the birefringent nature of the samples.

The birefringent origin of the observed optical rotation is further confirmed by the non-reciprocal ORD responses, **figure 9a-d**. The line shapes of the reflectance spectra are independent of the direction of propagation of the incident light. In contrast, the ORD resonances undergo both a change in sign and line shape when the sample is flipped around. This ORD behaviour is inconsistent with optical activity, which is reciprocal[27], and therefore must be dominated by the birefringence of the sample.

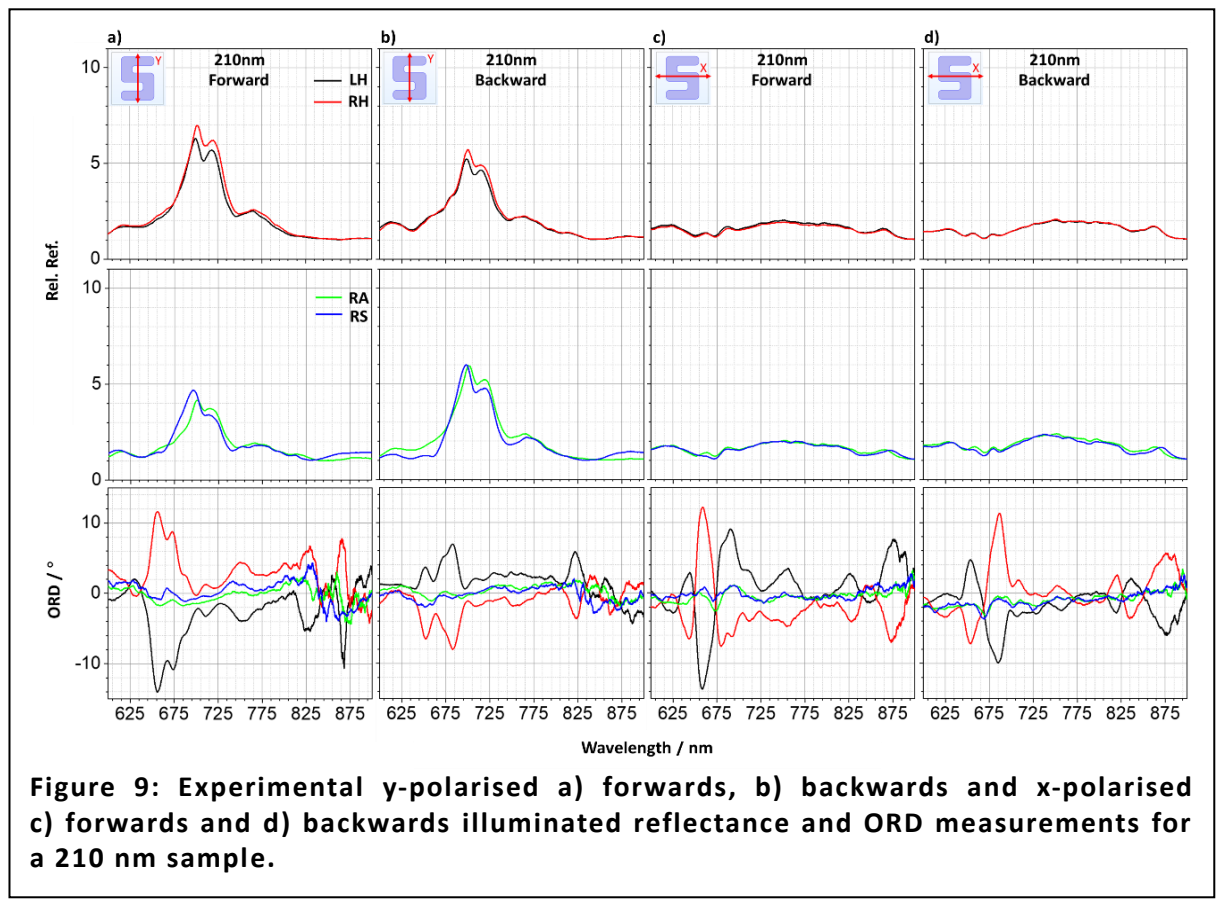

**Figure 9:** Experimental y-polarised a) forwards, b) backwards and x-polarised c) forwards and d) backwards illuminated reflectance and ORD measurements for a 210 nm sample.

## Mueller Matrix Polarimetry

Stokes polarimetry, which is predominately used to study chiroptical responses of metamaterials, cannot decouple chiroptical and birefringent effects. This requires a complete characterisation of a sample's interaction with light using Mueller Matrix Polarimetry (MMP).[28] Consequently, we have used MMP to collect circular dichroism / birefringence (CD/CB), linear dichroism / birefringence (LD/LB) and linear dichroism / birefringence at ±45° (LD'/LB') from 210 nm thick samples (see supp. **S8-10**). The MMP setup is housed at beamline B23 at Diamond Light Source Ltd, and is described elsewhere.[29] For brevity, only CD, LB, LD' and a simulated CD spectra are shown in **figure 10**.

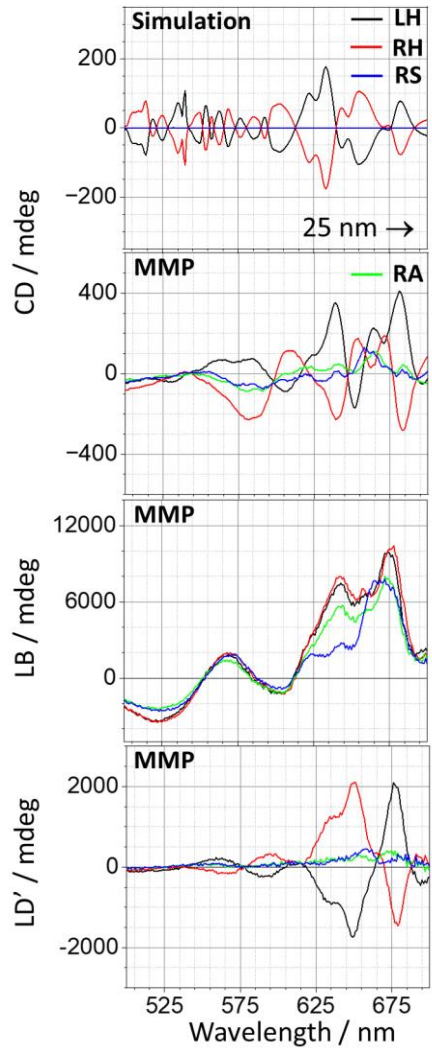

Figure 10: A comparison between simulated CD and MMP measured CD, LB and LD' for a 210 nm sample. The MMP CD data has been smoothed to aid comparison.

As would be expected based on the Stokes polarimetry data, the birefringent response is the dominant contribution, being ~ one order of magnitude larger than the optical activity measured by CD.

The CD spectra collected from LH and RH metasurfaces are approximately equal and opposite, with the RA and RS spectra consistently at the midpoint between the two. The simulated CD spectra is red shifted by 25 nm to account for the typical red shift associated with the idealised models, to aid comparison with experiment. There is reasonable agreement between simulated and experimental data, albeit that as expected, the former is more resolved. Experimentally, between 645-675 nm the spectra lie on an underlying background, which mimics relatively closely the CD response from RA and RS arrays. The LH/RH spectra would otherwise be expected to cross 0 mdeg at 675nm. When this is taken into account the simulation and experiment spectra are in good agreement. At lower wavelengths, the broad experimental peak across 575 nm appears as an oscillating feature in the simulation, which shows a positive (negative) mdeg bias for the LH (RH) array. Using a curve smoothing function over this wavelength range on the simulated spectra, which previously has been used to mimic the effects of structural heterogeneity, would improve the level of agreement with experiment.[30]

Although the spectral line shapes differ, the resonances associated with optical activity and birefringence do occur in similar positions. The most intense features occur within the 625 – 685 nm wavelength range in which the $E_Q$ and $M_D$ are located. The anisotropic properties of these modes are the origins of the birefringent resonances, however, optical activity has different origins. For isotropic chiral media, optical activity stems from the $E_D \cdot M_D$ cross term, which has a small magnitude and is the cause of the inherent weakness of chiroptical responses in solution phase. For anisotropic media $E_D \cdot E_Q$ is non-zero and can make a significant contribution to optical activity, particularly in the case of metamaterials where large field gradients enhance its influence.[31,32] Consequently, increasing the thickness of the silicon structures alters the optical activity because it shifts both the $E_D$ and $E_Q$ modes and hence changes the magnitude of the cross-terms.

## Optical Chirality

The chiral asymmetry of near fields can be parameterised using the optical chirality parameter ($C$)[33], which are typically normalised to the value of that of a helicity of circularly polarised light (in this current study, right-circularly polarised light (RCPL)). Near fields generated by scattering of even LPL from chiral nanostructures can generate near fields which in localised regions of space have significantly higher chiral asymmetries than circularly polarised light. This property is sometimes referred to as superchirality.[2,34] The levels of optical chirality of near fields generated by x- and y-polarised light as a function of silicon thickness have been derived from numerical simulations. **Figure 11** shows the volume averaged optical chirality calculated from the structure and surrounding volume above the substrate.

As expected of enantiomorphic structures, the optical chirality is equal and opposite between the LH and RH forms, the net contributions of which cancel in the case of the RS array. A notable aspect of the optical chirality is that the sign is wavelength dependent and is not solely determined by the handedness of the structure. Peaks in optical chirality corelate to wavelength regions where one would expect to find $E_D$, $M_D$ and $E_Q$ resonances. Indeed, the largest peaks in optical chirality are observed for y-polarised incident light which produces stronger $E_D$, $M_d$ and $E_Q$ resonances, than the equivalent ones for x-polarised illumination.

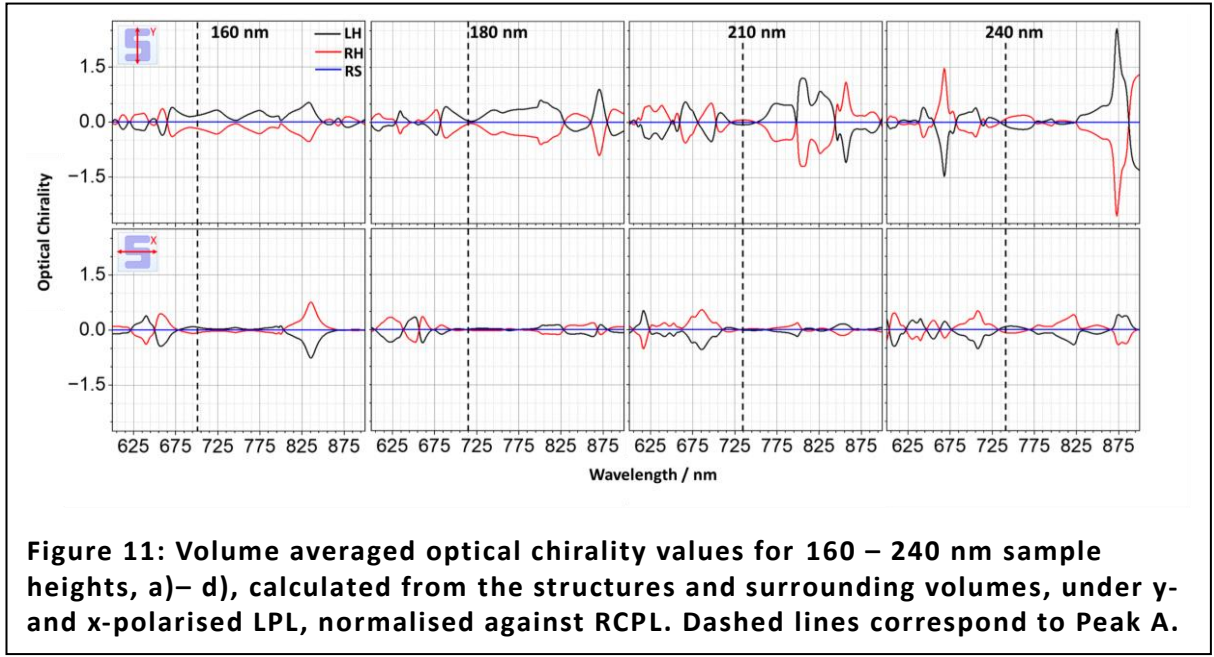

Figure 11: Volume averaged optical chirality values for 160 – 240 nm sample heights, a)– d), calculated from the structures and surrounding volumes, under y- and x-polarised LPL, normalised against RCPL. Dashed lines correspond to Peak A.

Maps shown in **figure 12** illustrate the spatial distribution of the chiral asymmetries of the near fields at wavelengths corresponding to peak A resonances, shown as dashed lines in the volume averaged optical chirality plot. These maps show equal and opposite behaviour expected of enantiomorphic structures. Fields within the silicon structures show high levels of chiral asymmetries (i.e., superchiral). However, there are relatively strong superchiral fields that exist near the surface of the structures and between adjacent arms, in the case of the y-polarised illumination; fields which could potentially be exploited for chiral sensing.

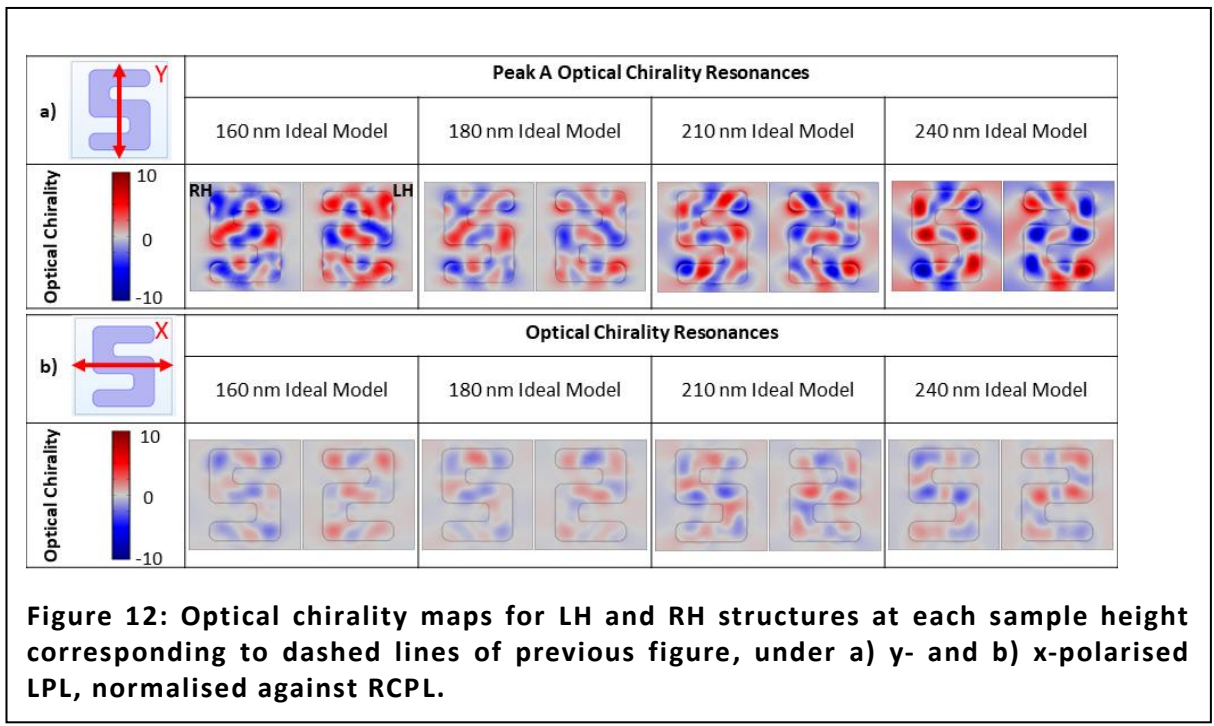

Figure 12: Optical chirality maps for LH and RH structures at each sample height corresponding to dashed lines of previous figure, under a) y- and b) x-polarised LPL, normalised against RCPL.

## Conclusions

We have presented chiral silicon metasurfaces which display strong optical rotation properties that are dependent on the height of the constituent nanostructures. The inherent birefringence of the structure is the dominant contribution to optical rotation, being an order of magnitude greater than the chirality derived optical activity. Using Mueller matrix polarimetry, the birefringent and optical activity contributions could be decoupled. The thickness dependence of the optical rotation could be understood through a multipole decomposition based on numerical simulations. The change in behaviour for the thickest sample was attributed to a combination of the resonance shifting as well as a large increase in magnetic quadrupole contribution. This change in multipolar character is facilitated by the greater ability of the thicker sample to sustain multipolar resonances that are supported across the height of the structures. The arrays were found to generate large enhancements of optical chirality in and around the nanostructures, which was also found to increase with sample height. In a broader context the present study has implications for the use of dielectric metasurfaces in applications such as enantiomeric/biomacromolecular detection. It illustrates the potential of birefringent effects to dominate the response from metasurfaces obtained using ubiquitous Stokes polarimetry methodologies. Non-reciprocity has often been falsely assumed to be the expected response from chiral planar metasurfaces formed from structures which in free space have horizontal mirror planes. Using MMP measurements clearly illustrates that such non-reciprocity originates from birefringence. Finally, using the multipolar analysis we illustrate how $E_D \cdot E_Q$ cross-term contributes both to the optical activity displayed by the structures and the enhanced chiral asymmetry of the near fields.

## Conflicts of Interest

There are no conflicts to declare.

## Author Contributions

The manuscript was written through contributions of all authors. All authors have given approval to the final version of the manuscript.

## Acknowledgements

The authors acknowledge financial support from the Engineering and Physical Sciences Research Council (EP/S012745/1 and EP/S029168/1). Technical support from the James Watt Nanofabrication Centre (JWNC) and Diamond Light Source Ltd. is acknowledged. DKM was awarded a studentship by the EPSRC. M.K. acknowledges the Leverhulme Trust for the award of a Research Fellowship (RF-2019-023).

## Notes

The authors declare no competing financial interests.

## Methodology

**Sample fabrication**

Samples were fabricated in the James Watt Nanofabrication Centre (JWNC). The s-structures were fabricated using an electron beam lithography process. Quartz glass slides were cleaned under ultrasonic agitation in acetone, methanol, and isopropyl alcohol (AMI) for 5 min each, dried under N2 flow and exposed to $O_2$ plasma for 5 min at 100 W. Amorphous silicon was deposited on the substrates through PECVD with a SPTS Delta tool. The samples were cleaned as before, with a 1 min lower power 60W plasma clean. A PMMA resist bilayer was then spun at 4000 rpm for 1 min and baked at 180 °C for 5 min between spins. A 10 nm aluminium conducting layer was evaporated on the substrates using a PLASSYS MEB 550s evaporator. Patterns were designed on the CAD software L-Edit and written by a Raith EBPG 5200 electron beam tool operating at 100 kV. The Al was removed in CD-26 and the resist developed in 3:1 MIBK: IPA solution at 23.2 °C for 1 min, rinsed in IPA for 5 seconds and then RO water before drying under $N_2$ flow. A 50 nm nichrome etch mask was then evaporated onto the sample prior to a lift-off procedure in acetone at 50 °C overnight, with the sample agitated to remove all remaining resist and excess metal. The silicon was etched using an STS tool with a custom recipe of $C_4F_8$/$SF_6$ 90/30 sccm, 600W, 9.8 mTorr, 20 °C for 3 minutes. The NiCr etch mask was removed by immersion in chromium etchant and 60% Nitric acid. The process was completed with a final AMI and low power plasma cleaning.

**Reflectance measurements**

Substrates were placed in a custom-printed sample holder, with a FastWell Silicone seal and clear borosilicate glass slide above it. The nanoarrays were immersed in PBS by injecting the buffer into the holder cavity. The sample was secured on the stage of a custom built stokes polarimeter. The incident polarisation was altered using a linear polariser and ORD was collected by measuring intensities at four angles of the analyser, which was rotated between 0°, 45°, 90° and 135°. Reflectance data was obtained with the analyser at 0°.

**Mueller Matrix Polarimetry**

Samples measured for MMP were prepared in the same manner as for the Stokes polarimeter. Two pairs of photoelastic modulators either side of the sample act as a polarisation state generator and polarisation state analyser, allowing for production and measurement of all light states required to obtain the 16 Mueller matrix elements.

**Numerical Simulations**

Simulations were carried out using commercial finite element analysis software, COMSOL Multiphysics v6.1 (Wave optics module). The nanostructure is surrounded by a cuboid representing a unit cell, with the x and y dimensions defining the periodicity of the metamaterial, as calculated from AFM images. The z dimensions of the cell are sufficiently large ($\geq \lambda_{max}/2$) that near-fields generated by the nanostructures do not extend to any integration surfaces above the structure, the total height of the cell is 1800 nm. The unit cell is split up into domains of varying thickness. The top and bottom 200 nm domains are perfectly matched layers (PMLs) which absorb all reflections from the nanoparticle. The boundary adjacent to the upper PML is the excitation port, from which incident light originates and its polarization is specified. Reflection intensity is also calculated at this port. The silicon structure is near the centre of the cuboid. The boundary adjacent to the lower PML is the outgoing port.